# The inverse Laplace transform as the ultimate tool for transverse mass spectra

Ekkard Schnedermann

Physics Department, Brookhaven National Laboratory

Upton, New York 11973, USA

20 Jan 1994

**Abstract**

New high statistics data from the second generation of ultrarelativistic heavy-ion experiments open up new possibilities in terms of data analysis. To fully utilize the potential we propose to analyze the $m_\perp$-spectra of hadrons using the inverse Laplace transform. The problems with its inherent ill-definedness can be overcome and several applications in other fields like biology, chemistry or optics have already shown its feasability. Moreover the method also promises to deliver upper bounds on the total information content of the spectra, which is of big importance for all other means of analysis. Here we compute several Laplace inversions from different thermal scenarios both analytically and numerically to test the efficiency of the method. Especially the case of a two component structure, related to a possible first order phase transition to a quark gluon plasma, is closer investigated and it is shown that at least a signal to noise ratio of $10^4$ is necessary to resolve two individual components.

## 1 Introduction

The hadronic transverse momentum spectra are one of the easiest accessible observables in relativistic heavy-ion collisions and are currently the main source of information about the dynamical evolution of the collision zone. With the arrival of new high statistics data from the second generation of heavy-ion experiments at CERN [2] and the planned high statistics experiments at RHIC (Relativistic Heavy Ion Collider) [3] new prospects for their analysis open up.

Customarily the analytical approach consists of performing a least squares fit of an exponential function to the measured $m_\perp$-spectra in order to determine the global slope, which might be directly or indirectly connected to the temperature of an equilibrated hadronic system. It has already shown its limitations in the sense that the current data cannot be described by a single exponential over the whole range in $m_\perp$ [4, 5]. Fitting instead a sum of two or even more exponentials (along with their normalizations) to the data becomes an increasingly ill-defined procedure as the possible errors become strongly correlated.

A straightforward extension of the method is to analyze the local slopes as a function of $m_\perp$ instead of the global slope. It can be performed for better theoretical insight into the spectral shapes from resonance decays or transverse flows, e.g. [6]. However, it seems to be impractical experimentally, because the approximation of the exponential slopes by finite differences between the measured data points will lead to huge error bars. Moreover, at least in our opinion as opposed to [7], the local slopes plotted over $m_\perp$ have no obvious physical meaning, thus making it difficult to interpret the slopes without resorting to an actual model calculation.

We want to propose the *inverse Laplace transform* for the analysis of almost exponentially decaying spectra like the $m_\perp$-spectra. It performs the analysis of a function in terms of decaying exponentials quite analogous to the Fourier transform, which employs oscillating exponentials. Though its use has been suggested several times in the past, the Laplace transform has not yet matched the tremendous success of the Fourier transform in data analysis; the numerous literature, e.g. [9], is rather focused on solving differential equations. Partly that is because the measurement has to be precise over several orders of magnitude, in analogy to several oscillations needed for the Fourier transform. But the main reason is that the construction of the inverse Laplace transform for real data is an inherently ill-defined or ill-conditioned problem [9, 11]. This problem has been solved by means of an eigenfunction or singular function analysis [12, 13] and many applications from areas as diverse as medicine, biology, chemistry and optics have been investigated [14], amongst them the retrieval of the temperature distribution inside the human body using microwave radiation.

After showing the problems with naive fit procedures we want to investigate in this paper the applicability of the inverse Laplace transform to ultrarelativistic heavy ion collisions. Assuming that a locally equilibrated system is produced in these collisions, particle spectra can be computed for a number of scenarios. Most of them can be inverted analytically to provide us with a set of patterns to look for in the inversion of real data. The actual inversion of data by means of a singular function analysis is then performed for several test-cases, focussing especially on the requirements for resolving a two component structure in the spectra.

## 2 Naive Methods

As a practical exercise we perform fits by sums of exponentials with $c_i$ and $t_i$ as the free parameters

$$g_{\text{fit}}(m_\perp) = \sum_{i=1}^{n} c_i \exp(-m_\perp t_i) \quad (1)$$

to a test spectrum, which we construct from two thermal distributions with different temperatures $T_1$ and $T_2$ integrated over rapidity (or fluid rapidity in a boost invariant situation):

$$g_{\text{test}}(m_\perp) = \frac{dn}{m_\perp dm_\perp} \propto \frac{m_\perp}{T_1} K_1\left(\frac{m_\perp}{T_1}\right) + \frac{m_\perp}{T_2} K_1\left(\frac{m_\perp}{T_2}\right) \quad (2)$$

where $m_\perp = \sqrt{m_0^2 + p_\perp^2}$ is the transverse mass of the particle.

| $c_1$ | $t_1 \times 200\,\text{MeV}$ | $c_2$ | $t_2 \times 200\,\text{MeV}$ | $c_3$ | $t_3 \times 200\,\text{MeV}$ |
|---|---|---|---|---|---|
| 1.06 | 0.87 | -0.30 | 1.47 | | |
| 1.12 | 0.87 | -0.26 | 1.26 | -0.10 | 1.88 |
| 1.09 | 0.87 | -0.30 | 1.36 | -0.04 | 2.37 |
| 1.08 | 0.87 | -0.31 | 1.41 | -0.02 | 3.16 |

**Table 1:** Fitting the spectrum (2) by a sum of exponentials (1) gives a stable fit in the case of two slopes with a good agreement ($\chi^2 \approx 2 \times 10^{-5}$) to the test function, which however does not reproduce the built in components at $t = 0.9/200\,\text{MeV}$ and $t = 1.1/200\,\text{MeV}$. A three slope fit does not improve this situation but is already difficult to perform due to the many local minima in the six dimensional parameter space from which we are showing the three fits with $\chi^2 \lesssim 10^{-6}$

For a specific example we have chosen $m_0 = 140\,\text{MeV}$, $T_1 = 200/0.9\,\text{MeV}$ and $T_2 = 200/1.1\,\text{MeV}$ so that the two components should appear at $t_1 = 0.9/200\,\text{MeV}^{-1}$ and $t_2 = 1.1/200\,\text{MeV}^{-1}$. Note that we adjusted the relative normalizations so that their Laplace components (12) are roughly equal in size, although the number of particles from the higher temperature dominate the ones from lower temperature by a factor of 1.56. The quality of the fits is then judged by summing the squared difference over 32 data points $p_i$ spaced at 50 MeV intervals:

$$\chi^2 = \sum_{i=1}^{32} \left(1 - \frac{g_{\text{fit}}(p_i)}{g_{\text{test}}(p_i)}\right)^2 \qquad (3)$$

| | $t_i \times 200$ MeV | | | | | | |
| | 0.8 | 0.9 | 1.0 | 1.1 | 1.2 | 1.3 | |
| $n_{\text{dim}}$ | $c_1$ | $c_2$ | $c_3$ | $c_4$ | $c_5$ | $c_6$ | $\chi^2$ |
| 2 | | 1.48 | | -0.70 | | | $4 \times 10^{-3}$ |
| 3 | | 1.33 | 0.44 | -0.99 | | | $2 \times 10^{-3}$ |
| 4 | 0.63 | -1.43 | 4.35 | -2.78 | | | $3 \times 10^{-4}$ |
| | | 2.32 | -3.87 | 5.03 | -2.72 | | $6 \times 10^{-5}$ |
| 5 | -0.49 | 5.22 | -10.10 | 10.85 | -4.72 | | $2 \times 10^{-5}$ |
| 6 | 0.84 | -4.65 | 18.63 | -30.21 | 24.11 | -7.96 | $1 \times 10^{-6}$ |

**Table 2:** Fits to the spectrum (2) by sums of exponentials (1) with the slopes $t_i$ fixed and the normalizations $c_i$ as the free parameters. The obvious instabilities in the solution grow with the number of parameters and render this method unusable for higher precision data.

We performed two different fit procedures. In the first case, shown in Tab. 1, we performed a fit to the spectrum (2) with both the $c_i$ and $t_i$ as free parameters. Because of the many local minima involved this becomes a rather tricky fit for more than three exponentials so that in the second case, shown in Tab. 2, we fixed the slopes $t_i$ at chosen values and only performed fits with the normalizations $c_i$ as free parameters.

From the study of the tables it becomes apparent that the fits are becoming more and more unstable when the number of parameters is increased and are thus meaningless. In no case were the fits able to capture the built-in two component nature of the spectrum, which we ascribe to the high sensitivity of this prescription to small "fluctuations" in combination with the fact that the Bessel functions are only approximately exponential. In the absence of noise and for pure exponentials there is apparently no problem, although the $\chi^2$ contours are elongated ellipsoids, thus showing very big correlated errors between the two slope parameters. In the following sections we will demonstrate that the inverse Laplace transform by the method of singular functions is much better suited to the occurrence of deviations from exponential behaviour and noise in the data.

## 3  Analytical Inversions

The Laplace transform of the function $f(t)$ is given by

$$g(p) = \mathcal{L}\{f\} = \int_0^\infty f(t)\, e^{-pt}\, dt \qquad (4)$$

and is a linear transformation. This has the effect that different Laplace components $f(t) = \sum f_i(t)$ with their corresponding spectra $g_i(p)$ will give a spectrum which again is the sum of its components $g(p) = \sum g_i(p)$. This seemingly trivial property is not valid for the analysis in terms of local slopes, so that the connection between a certain slope value and the underlying physics is rather hidden.

In the case of an extended static source in local thermal equilibrium we would interpret the Laplace components $f(t)$ as the temperature distribution (actually $1/T$-distribution) of the source. For hadronic spectra we would expect that they are concentrated around the freeze-out temperature, where the equilibrium between particles can no longer be kept up and they are streaming freely into the detectors. The appearance of a two component structure in the inverse $f(t)$ would hint towards the presence of a phase transition into the quark gluon plasma. The second peak, located at $t = 1/T_c$, could originate from the hadrons which escape from the mixed phase during its long life time spent at the critical temperature $T_c$, and travel to the detectors without further interactions in the surrounding hadron gas.

To obtain the components $f(t)$ from the measured spectrum $g(p)$, or $g(m_\perp)$, we have to perform the inverse Laplace transformation. It is given by Bromwich's integral in the complex plane

$$f(t) = \mathcal{L}^{-1}\{g\} = \frac{1}{2\pi i} \int_{c-i\infty}^{c+i\infty} g(p)\, e^{-pt}\, dp \qquad (5)$$

with $c$ any real number so that $g(p)$ is analytical for $\operatorname{Re} p \geq c$. We will use this method, or a numerical implementation of it via the NAG-Library, to compute the inverse Laplace transforms for several spectral shapes which follow from theoretical scenarios. However, for measured data a different procedure has to be developed in section 4, because they are not available in the complex plane.

Nevertheless, inversions using eq. (5) can be performed for many theoretical scenarios, if the theoretically computed spectra have unique extensions into the complex plane. We give inverse transforms here for several kinds of (hadron) $m_\perp$-spectra, because we think they have the best chances of application, though also $p_\perp$-spectra and (dilepton) invariant mass spectra might be conceivable. We list several formula (with the help of [10]), which relate to transforming the spectrum shifted by the restmass (7), multiplication by $m_\perp$ (8), Boltzmann distributions (10), integrated thermal distributions (12) and Bose/Fermi-distributions (13).

$$\mathcal{L}^{-1}\{g(m_\perp)\} = f(t) \tag{6}$$

$$\mathcal{L}^{-1}\{g(m_\perp)\} = \exp(-tm_0)\mathcal{L}^{-1}\{g(m_\perp - m_0)\} \tag{7}$$

$$\mathcal{L}^{-1}\{m_\perp g(m_\perp)\} = \frac{d}{dt}\mathcal{L}^{-1}\{g(m_\perp)\} - \lim_{t\to 0+}\mathcal{L}^{-1}\{g(m_\perp)\} \tag{8}$$

$$\mathcal{L}^{-1}\{\exp(-m_\perp/T)\} = \delta(t - \frac{1}{T}) \tag{9}$$

$$\mathcal{L}^{-1}\{m_\perp \exp(-m_\perp/T)\} = \delta'(t - \frac{1}{T}) \tag{10}$$

$$\mathcal{L}^{-1}\{K_1(m_\perp/T)\} = \begin{cases} 0 & \text{for } 0 \leq t \leq \frac{1}{T} \\ \frac{tT^2}{\sqrt{(tT)^2-1}} & \text{for } t > \frac{1}{T} \end{cases} \tag{11}$$

$$\mathcal{L}^{-1}\{m_\perp K_1(m_\perp/T)\} = \begin{cases} 0 & \text{for } 0 \leq t \leq \frac{1}{T} \\ \delta(t - \frac{1}{T}) - \frac{T^2}{\sqrt{(tT)^2-1}^3} & \text{for } t > \frac{1}{T} \end{cases} \tag{12}$$

$$\mathcal{L}^{-1}\{\frac{m_\perp}{\exp(m_\perp/T) \mp 1}\} = \sum_{n=1}^{\infty}(\pm)^{n-1}\delta'\left(t - \frac{n}{T}\right) \tag{13}$$

The spectral shape of (12) involving the Bessel function $K_1$ results from integrating the thermal spectrum of a point source over rapidity to compute $dn/dm_\perp^2$, or equivalently, to integrate over the fluid rapidity $\eta$ in a boost invariant scenario to compute $dn/dy\, dm_\perp^2$ (e.g. [6]).

## 4 Results

In the case of real data, which are naturally given only on the real axis, we have to proceed differently, because inverting the Laplace transform for functions which are defined only on the real axis instead of at least a complex half plane is an ill-conditioned problem [11]. For data points with associated measurement errors the problem even gets worse. Historically there have been many naive attempts to compute the inverse Laplace transform, e.g. by fitting the data with superpositions of exponentials (e.g. [8])

$$f(t) = \sum_{i=1}^{n} c_i \, \delta(t - t_i), \qquad (14)$$

or combinations of exponentials and polynomials [15].

This is not a viable method because the inverse Laplace transform is not continuous if no further restrictions on the possible space of ordinary functions are imposed. This can be illustrated by the following Laplace transform [9]:

$$\mathcal{L}\{\sin \omega t\} = \frac{\omega}{p^2 + \omega^2} \qquad (15)$$

which can be made arbitrarily small with $\omega \to \infty$. One could thus approach with $\mathcal{L}f$ the given data while $f$ is not at all converging to a certain solution. This can be cured by expanding $f$ in terms of a series of singular functions [11, 12, 13]. This effectively restricts the available functional space for $f$ by cutting out high frequency oscillations (=noise), which would otherwise drastically influence the solution.

We follow the method given in [13], which assumes that $f$ is roughly distributed like some profile function $P(t)$, which is in this case chosen to be the Gamma distribution with width $\beta$:

$$P(t) = \frac{\beta^\beta}{\Gamma(\beta)} t^{\beta-1} \exp(-\beta t) \qquad (16)$$

This profile function can be tailored to the specific problem at hand, even step functions can be chosen [12]. The width $\beta$ can be taken, as we did, from pure prejudice about what the components should be. The choice can actually be checked, because if the window profile is chosen too small the inversion will try to fold the outside components into the window which will in turn lead to big oscillations and will depend strongly on the actual number of singular values used. However, ideally the width would have to be estimated from the data, e.g. by the cumulants method [15], which is however a rather complicated procedure in itself.

In terms of data sampling we are not as diligently adjusting the spacing of the data points as it is done in [13] in order to retrieve maximal information content. We simply assume that the data consists of 32 points taken at 50 MeV intervals, though apparently exponential sampling [16] is more efficient, especially when it is matched to the width of the profile function. However, our choices are not very far from optimal because our necessary signal to noise ratio seems to be in line with [13].

Because there is only a finite number of sampling points, the Laplace inversion by means of singular functions turns into a singular value decomposition for ordinary matrices, which can easily be performed numerically and which yields a descending series of *singular values* $\alpha_i$. By truncating that series at a certain $\alpha_n$, the part of the solution which leads to wild oscillations is filtered out and one obtains the approximate solution as a series of *singular functions*. The precision of the regularized solution, i.e. the noise to signal ratio, is reflected in the condition number $\alpha_n/\alpha_1$. Naturally, also more elaborate regularization schemes can be used [17].

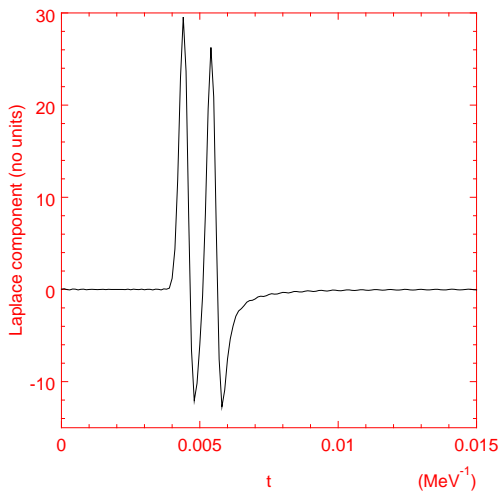 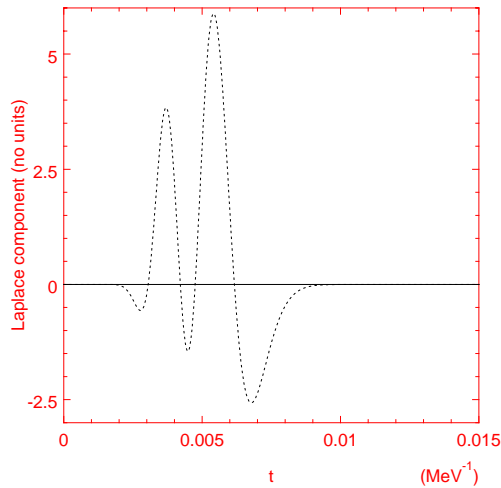

**Figure 1:** Numerical inversion of the spectral function as performed by the NAG-Routines. We assumed a spread in the thermal components of the size of 5%. The particle's mass has been fixed as the pion mass.

**Figure 2:** Reconstruction of two thermal pion distributions with $T_f = 180\,\mathrm{MeV}$ and $T_c = 220\,\mathrm{MeV}$ using 32 data points spaced at 50 MeV and a very narrow profile function with $\beta = 20$. A clear seperation of the two components is seen for the case of 5 singular functions, corresponding to a noise level of $\alpha_5/\alpha_1 \lesssim 1.5 \times 10^{-4}$.

Some general examples of inversions are shown in [13]. Here we will concentrate on several cases which are more closely related to possible heavy-ion situation scenarios and we return to our initial motivation, which is to search for the phase transition by looking for a two component structure. In Figs. 1 and 2 we investigate the resolution limits of the inversion method for two thermal distributions which are very close together (2), i.e. $T_f = 182\,\text{MeV}$ for the freeze-out component and $T_c = 222\,\text{MeV}$ for the critical temperature of the phase transition. Especially with respect to $T_c$ those values are meant only as moderately realistic examples which are modeled after the currently measured spectra. Note that the apparent temperatures do not have to be the actual local temperatures in the system since radial flows of average $\langle v_r \rangle \approx 0.3\,c$ give rise to a blueshift of about 30% [6].

First, in Fig. 1, we have used the NAG-Library to perform a Laplace inversion of a two-component thermal spectrum of pions, which has been extended to the complex plane, to have a benchmark for the following inversion of data on the real axis. In Fig. 2 we then perform a Laplace inversion from a singular value decomposition, chosing a very narrow profile function ($\beta = 20$) centered at $t = 1/200\,\text{MeV}^{-1}$. We see that we indeed can seperate both components much better than with the naive attempts section 2, though the reproduction of the shape of the actual inverse Laplace transform is rather marginal with this noise level. Nevertheless, this scenario shows the basic applicability of the method to actual heavy-ion data and a promising new way to analyze high statistics data.

However, we note that the actual heavy-ion collision might produce apart from the purely thermal radiation also resonances which afterwards decay into pions [18]. Again to enable comparison with Fig. 4 we perform in Fig. 3 the inversion of an integrated thermal distribution (12) of pions at $T = 200\,\text{MeV}$ using a profile function centered at $T$ with a width $\beta = 2$, which is much broader than the actual distribution. We note that the reconstruction, while much broader than the actual inversion, is still narrower than the profile function. Moreover we do not attribute any physical meaning to the fluctuations at low $t$, which might simply reflect an unfortunate choice of profile function $P(t)$.

In Fig. 4 the additional pions from decays of a thermal population of $\rho$ and $\omega$ resonances (modeled after [18]) clearly gives rise to a second component which is well seperated from the first one. Its appearance at larger values of $t$ is linked to the much lower apparent temperatures $T_{\text{app}} = 1/t$ of resonance decay spectra when compared to the actual temperature of the system [18, 6]. The inversion requires at least five singular functions (shown is seven) and a noise to signal ratio of at least $\alpha_5/\alpha_1 \lesssim 5 \times 10^{-3}$ ($\alpha_7/\alpha_1 \lesssim 2 \times 10^{-4}$).

Apparently other processes, e.g. resonance decays, can also simulate a second temperature component, which is a drawback to the proposed method.

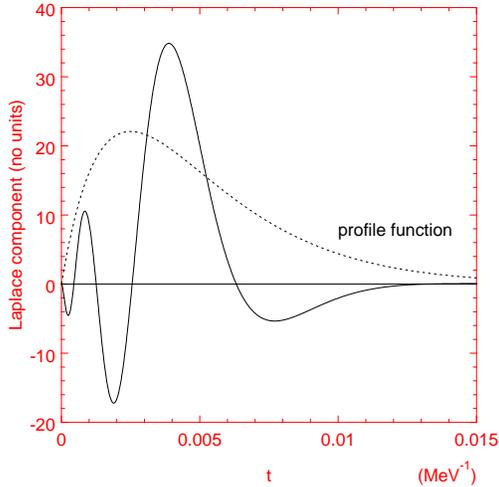 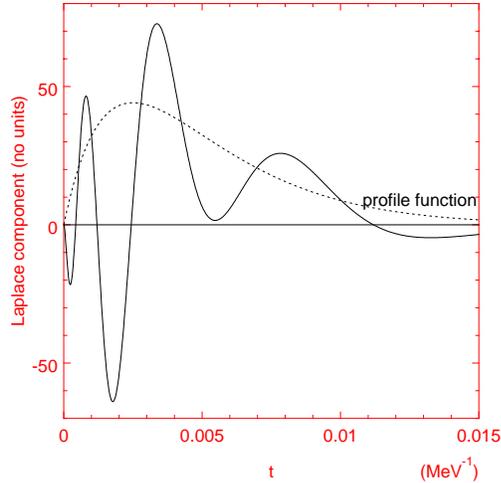

**Figure 3:** Laplace inversion of a thermal pion $m_\perp$-spectrum using 32 datapoints spaced at 50 MeV and using the first seven singular values $\alpha_i = 0.71, 0.20, 0.059, 0.015, 0.0036, 0.00081$ and $0.00017$. The profile $P(t)$ was chosen very wide ($\beta = 2$).

**Figure 4:** The same Laplace inversion as in Fig. 3 now for a pion spectrum additionally including $\rho$ and $\omega$ decays shows a second component at higher $t$. This reconstruction corresponds to a noise level of $\alpha_7/\alpha_1 \lesssim 2 \times 10^{-4}$.

Nevertheless, there are two possibilities around this obstacle, because the decay contributions necessarily depend on the kind of particle measured and the produced resonances. By selecting particle types which are less sensitive to the resonance contribution than the pions, e.g. kaons, and comparing them among one another, one could eliminate large parts of that background, because only the thermal components should look similar across all particle species. On the other hand, if there are very many resonance channels, the multitude of channels might rather form a continuum of Laplace components, above which certain peaking thermal components at $T_f$ and $T_c$ might still be identifiable.

## 5  Summary and Conclusion

We have investigated the applicability of the inverse Laplace transform on heavy-ion data and have shown that naive methods like fitting multiple exponentials are incapable to analyze high statistics data. Using the method of singular function decomposition [13] we have shown how a two component structure in the produced hadron spectra can be identified and that the re-quirements for the data in terms of noise level are within reach of current or planned experiments. We also would like to argue that the issue of systematical errors becomes less severe because the method projects some errors into certain regions so that they are eliminated or identifyable. However, only a simulation of the actual experiments could assure that.

However we have also shown how resonance decays can cloud the picture by generating a second component at about twice the thermal slope. Performing an analysis with kaons or baryons should reduce this component and also change its shape so that the whole body of data should give clear answer.

We have tried a Laplace inversion with $\pi^+$ and $\pi^-$ data from the ISR [19] which have statistical errors of about 1-2%. However, we could not find any indication of a two component structure, which might either indicate the absence of a (first order) phase transition or insufficient data to decide the question.

The hadron spectra are the most promising for an analysis because of their high statistics. The inversion of the lepton spectra might be even more interesting, e.g. one would look for a single component at $T_c$ above a rather smooth background, but the necessary precision seems to be far out of reach.

Further improvements of the singular function method, e.g. exponential sampling, choice of profile function, can conceivably improve upon the efficiency of the method, but they will require a close relation to the performed experiment.

# 6 Acknowledgments

E. S. gratefully acknowledges support by the Alexander-von-Humboldt Stiftung and the U.S. Department of Energy under contract numbers DE-AC02-76H00016 and DE-FG02-93ER40768.